\documentclass[nonacm,sigconf]{acmart}

\AtBeginDocument{%
  }

\acmConference[SPAA]{ACM Symposium on Parallelism in Algorithms and Architectures}{June 17—21, 2024}{Nantes, France}

\usepackage{algorithmic}
\usepackage{caption}
\usepackage[linesnumbered,lined,boxed,commentsnumbered,ruled,longend, vlined]{algorithm2e}

\newcommand{\T}{{\mathcal T}}
\newcommand{\Tl}{\textit{et al.}}

\newtheorem{theorem}{Theorem}

\newtheorem{lemma}{Lemma}

\newtheorem{example}{Example}
\usepackage{enumitem}
\usepackage{mathtools}
\usepackage{todonotes}
\newcommand{\dk}[1]{{\color{black}#1}}
\newcommand{\ra}[1]{{\color{black}#1}}

\copyrightyear{2024} 
\acmYear{2024} 
\setcopyright{rightsretained} 
\acmConference[SPAA '24]{Proceedings of the 36th ACM Symposium on Parallelism in Algorithms and Architectures}{June 17--21, 2024}{Nantes, France}
\acmBooktitle{Proceedings of the 36th ACM Symposium on Parallelism in Algorithms and Architectures (SPAA '24), June 17--21, 2024, Nantes, France}\acmDOI{10.1145/3626183.3659970}
\acmISBN{979-8-4007-0416-1/24/06}

\begin{document}

\title{Stable Blockchain Sharding under Adversarial
Transaction Generation}



\author{Ramesh Adhikari}
\orcid{0000-0002-8200-9046}
\affiliation{%
  \institution{School of Computer \& Cyber Sciences\\Augusta University}
  \city{Augusta}
  \state{Georgia}
  \country{USA}
  \postcode{30912}
  }
\email{radhikari@augusta.edu}

\author{Costas Busch}
\orcid{0000-0002-4381-4333}
\affiliation{%
  \institution{School of Computer \& Cyber Sciences\\Augusta University}
  \city{Augusta}
  \state{Georgia}
  \country{USA}
  \postcode{30912}
}
\email{kbusch@augusta.edu}

\author{Dariusz R. Kowalski}
\orcid{0000-0002-1316-7788}
\affiliation{%
  \institution{School of Computer \& Cyber Sciences \\ Augusta University}
  \city{Augusta}
  \state{Georgia}
  \country{USA}
  \postcode{30912}
}
\email{dkowalski@augusta.edu}









\begin{abstract}
Sharding is used to improve the scalability and performance of blockchain systems.
We investigate the stability of blockchain sharding, 
where transactions are continuously generated 
by an adversarial model.
The system consists of $n$ processing nodes that are divided into $s$ shards.
Following the paradigm of classical adversarial queuing theory, 
transactions are continuously received at injection rate $\rho \leq 1$ and burstiness $b > 0$.
We give an absolute upper bound $\max\{ \frac{2}{k+1},  \frac{2}{ \left\lfloor\sqrt{2s}\right\rfloor}\}$ on the maximum injection rate 
for which any scheduler could guarantee bounded queues and latency of transactions,
where $k$ is the number of shards that each transaction accesses.
We next give a basic \dk{distributed} scheduling algorithm for uniform systems where shards are equally close to each other.
To guarantee stability, the injection rate is limited to $\rho \leq \max\{ \frac{1}{18k}, \frac{1}{\lceil 18 \sqrt{s} \rceil} \}$.
We then provide a fully distributed scheduling algorithm for non-uniform systems where shards are arbitrarily far from each other.
By using a hierarchical clustering of the shards,
stability is guaranteed with injection rate 
$\rho \leq \frac{1}{c_1d \log^2 s} \cdot \max\{ \frac{1}{k}, \frac{1}{\sqrt{s}} \}$,
where $d$ is the worst distance of any transaction to the shards it will access, and $c_1$ is some positive constant.
We also conduct simulations to evaluate the algorithms and measure the average queue sizes and latency throughout the system.
To our knowledge, this is the first adversarial stability analysis of sharded blockchain systems.
\end{abstract}



\maketitle

\section{Introduction}
\label{sec:introduction}
A blockchain is a chain (linked list) of transaction blocks. Due to its several special features, such as fault tolerance, transparency, non-repudiation, and immutability \cite{survey-of-onsensus}, it has been used in various domains, such as cryptocurrency \cite{bitcoin, ethereum}, healthcare \cite{mcghin2019blockchain}, digital forensics \cite{akbarfam2023forensiblock, akbarfam2023deep}, and supply chain management \cite{azzi2019power}. In a blockchain network, all of the participating nodes need to reach a consensus to append a new block, which is a time and energy-consuming process. Moreover, each node is required to process and store all transactions, which leads to scalability issues in the blockchain system. 
To improve scalability and performance, {\em sharding} protocols have been proposed, such as Elastico \cite{Elastico}, OmniLedger \cite{OmniLedger}, RapidChain \cite{Rapidchain}, and ByShard \cite{Byshard}. Sharding divides the blockchain system into multiple shards, where each shard is a cluster of nodes.
The shards allow to process transactions in parallel. 
 
We propose to investigate the stability of sharded blockchain systems in which transactions are continuously generated by an adversarial model.
The arrival of the transactions in the system 
depends on the transaction generation rate 
$\rho$ (also called injection rate), which is the number of 
generated transactions amortized per time unit, 
and burstiness $b$, which models spontaneous arrivals of transactions not predicted by injection rate. 
The goal is to design distributed scheduling algorithms for the sharded blockchains  
so that the number of pending transactions is bounded (bounded queues) and the amount of waiting time for each transaction is small (low latency). 
Stability in blockchains is important for scalability and to improve the resiliency to Denial of Service attacks
where malicious nodes try to inject bursts of transactions
into the system in order to delay other transactions.

The adversarial model that we use for generating transactions in blockchain sharding is motivated by the adversarial queuing theory introduced by Borodin $\Tl$ \cite{borodin2001adversarial}. This theory has been used to analyze the stability of routing algorithms with continuous packet injection into a network 
\cite{chlebus2009maximum,chlebus2012adversarial}. 
In communication networks, the transmission of data packets depends on constraints determined by the network's characteristics, such as its topology and the capacity of its links or channels. Similarly, in the context of blockchain sharding, executing multiple transactions concurrently across the different shards is constrained by the property that each transaction must require exclusive access to every account it intends to interact with, which prevents the execution of several transactions at once.

We consider the system consisting of $n$ nodes, which are further divided into $s$ shards.
Each shard is responsible for handling a subset of the accounts.
A transaction $T$ is generated at one of the shards, which is the {\em home shard} for $T$. 
Similar to other sharding systems~\cite{Byshard, adhikari2023lockless}, each transaction $T$ is split into subtransactions,
where each subtransaction accesses an account. 
A subtransaction of $T$ is sent to the {\em destination shard} that holds the respective account.
Each destination shard maintains a local blockchain of the subtransactions that are sent to it.
The whole blockchain can be recreated by taking the union of the 
local blockchains at the shards~\cite{adhikari2023lockless}.

Each home shard has an {\em injection queue} 
that stores the pending transactions to be processed.
The home shard picks transactions from its injection queue
and sends their subtransactions to the 
respective destination shards. 
All home shards process transactions concurrently.
A complication arises when they pick {\em conflicting} transactions
that access the same account.
In such a case, the conflict prohibits the transactions to commit concurrently
and forces them to serialize.
A {\em scheduling algorithm}
coordinates the home shards and destination shards to process the transactions (and respective subtransactions) in a conflict-free manner.
The main performance metric for the scheduler is its ability to handle the maximum transaction generation rate while maintaining system stability within adversary constraints. Additional performance metrics include the size of the pending transaction queues and transaction latency. 

 \paragraph{\bf Contributions}
To our knowledge, we give the first comprehensive adversarial queuing theory analysis for blockchain sharding systems. 
We provide the following contributions:
\begin{itemize}[leftmargin = 1.5em]

\item 
\dk{{\em Injection Rate Limit:} We prove an upper bound on injection rate $\rho \leq \max\{ \frac{2}{k+1},  \frac{2}{ \left\lfloor\sqrt{2s}\right\rfloor}\}$ where $k$ is the number of shards that each transaction accesses, for which a stable scheduling algorithm is feasible.}

    \item {\em Basic Scheduling Algorithm for Uniform Model:} In the uniform communication model, any shard can communicate with any other shard within a single round. The uniform model is appropriate for systems that have strict guarantees for communication delay (e.g. multiprocessor systems). The algorithm runs in epochs, and in each epoch, one of the shards acts as {\em leader shard}, which receives the transaction information from all the home shards. 
    Then the leader shard calculates a schedule by coloring the conflict graph of transactions. The schedule is communicated with the home and destination shards. This algorithm can process transactions with a generation rate limited to $\rho \leq \max\{ \frac{1}{18k}, \frac{1}{\lceil 18 \sqrt{s} \rceil} \}$.
    Moreover, we prove that the number of pending transactions at any round is at most $4bs$ (which is the upper bound on queue size in each shard), and the latency of transactions is bounded by $36b \cdot \min \{k, \lceil \sqrt{s} \rceil \}$.
    
    \item {\em Fully Distributed Scheduling Algorithm:} We introduce a distributed transaction scheduling algorithm designed to schedule transactions in a decentralized manner without requiring a central authority. Moreover, this algorithm works for the non-uniform communication model. The algorithm is based on a hierarchical clustering of the shards. This scheduler remains stable for a transaction generation rate 
    $\rho \leq \frac{1}{c_1d \log^2 s} \cdot \max\{ \frac{1}{k}, \frac{1}{\sqrt{s}} \}$,
    where $d$ is the maximum distance for any transaction between its home shard and the destination shards it will access, and $c_1$ is some positive constant.
    For this scheduling algorithm, we also provide the upper bound on queue size as $4bs$ and transaction latency is at most 
        $2 \cdot c_1 bd \log^2 s \cdot min \{k, \lceil \sqrt{s}\rceil\}$.

    \item {\em Simulation Results:} To evaluate the performance of our proposed algorithms, we conducted simulations to measure the average queue size of pending transactions and transaction latency.
\end{itemize}

\paragraph*{\bf Paper Organization}
The rest of this paper is structured as follows: Section \ref{sec:related-work} provides related works. Section \ref{preliminaries} describes the preliminaries for this study and the sharding model.  We prove the upper bound on a stable injection rate in Section \ref{sec:lower-bound} .
Section~\ref{sec:basic-distributed-scheduler} presents a basic stable solution.
Section~\ref{sec:fully-distributed} generalizes the techniques to a fully distributed setting.
In Section \ref{sec:simulation-result}, we provide the simulation results.  
Finally, we give our discussion and conclusions in Section~\ref{sec:conclusion}.

\section{Related Work}
\label{sec:related-work}
    In the field of blockchain research, various proposals have been presented to tackle scalability challenges in the consensus layer~\cite{Jalal-Window,jalalzai2019proteus,jalalzai2021hermes}. Although these protocols have made some progress in improving scalability, the system's performance still suffers as the network size expands. Blockchain protocols must also guarantee transaction safety, as defined by ACID properties \cite{transaction-processing}. To address the blockchain scalability issue, several sharding protocols have been proposed, such as Elastico \cite{Elastico}, Rapidchain \cite{Rapidchain}, OmniLedger \cite{OmniLedger}, Byshard \cite{Byshard}, SharPer \cite{amiri2021sharper}, Lockless blockchain sharding with multi-version control \cite{adhikari2023lockless}, and the work \cite{dang2019towards}. These protocols have shown promising enhancements in transaction throughput. 
However, none of these protocols have specifically explored stable transaction scheduling techniques with sharding.

Extensive research has been conducted on transaction scheduling in shared memory multi-core systems as well as distributed systems. A recent work \cite{busch2023stable} introduced a stable scheduling algorithm designed specifically for software transactional memory systems under adversarial transaction generation. Moreover, transaction scheduling in distributed systems has been explored 
and aimed to achieve provable performance bounds on minimum communication cost ~\cite{attiya2015directory, sharma2014distributed,sharma2015load}. However, these works do not address transaction scheduling problems in the context of blockchain sharding. The main reason that the results ~\cite{attiya2015directory, sharma2014distributed,sharma2015load} cannot apply to blockchain sharding is the mobility of the objects. In their transactional memory models, an object can move from one node to another node where the transactions that request it reside. In our blockchain sharding model, the objects have fixed positions in their respective shards.

Adversarial queuing theory was proposed by Borodin \Tl~\cite{borodin2001adversarial}, which has been applied to study the stability of routing algorithms with continuous packet injection into a network. 
More generally, this is a technique that measures the stability of processing incoming data (i.e. transactions) without making any statistical assumptions about data generation.
In the dynamic environment, the adversarial queuing theory provides a framework for establishing worst-case performance bounds for deterministic distributed algorithms. Moreover, this theory has been applied to different dynamic tasks in communication networks. The work in \cite{bender2005adversarial} examines the worst-case performance of randomized backoff on simple multiple-access channels. Similarly, in an adversarial environment, \cite{chlebus2009maximum} provides maximum throughput of multiple access channels, where their protocol achieves throughput 1 for any number of stations against leaky-bucket adversaries. Moreover, \cite{chlebus2012adversarial} investigates deterministic distributed broadcasting in multiple~access~channels. 

As adversarial queuing theory has been applied in different dynamic task generations, so we are relating this theory to blockchain sharding, where the sharding system is unaware of the number of transactions generated in any time interval.

\section{Technical Preliminaries}
\label{preliminaries}

\paragraph{\bf Blocks and Blockchain}
A blockchain is a decentralized peer-to-peer ledger replicated across multiple interconnected nodes.
A blockchain is implemented as a linked list (chain) of blocks where each block consists of a sequence of transactions.  
Blocks are linked through hashes, which makes them immutable. Blocks are appended to the blockchain with a distributed consensus mechanism.

\paragraph{\bf Blockchain Sharding}
Similar to previous works~\cite{adhikari2023lockless,Byshard}, we consider a blockchain system with $n$ nodes 
partitioned into $s$ shards $S_1, S_2,\dots, S_s$ such that $S_i \subseteq \{1, \ldots, n\}$, for $i \neq j$, $S_i \cap S_j = \emptyset$, and $n = \sum_i |S_i|$.
Let $n_i = |S_i|$ denote the number of nodes in shard $S_i$. 


Shards communicate with each other through message passing.
\ra{In this paper, we are not focusing on optimizing the message size. In the worst case, the message size in our model is upper-bounded by $O(bs)$.}
Moreover, all non-faulty nodes in a shard agree on each message before transmission
(e.g. using PBFT within the shard \cite{PBFT}). Similar to previous work~\cite{Byshard,hellings2022fault}, we 
assume that we are given a
{\em cluster-sending} protocol for reliable and secure communication between shards, satisfying the following properties 
for transmitting data $\Re$ from shard $S_i$ to shard $S_j$:
    (1) Shard $S_i$ sends $\Re$ to $S_j$ if there is an agreement among the non-faulty nodes in $S_i$ to send $\Re$;
    (2) All non-faulty nodes in recipient shard $S_j$ will receive the same data $\Re$;
    (3) All honest nodes in sender shard $S_i$ receive confirmation of data $\Re$ receipt.
    \ra{We assume that these properties are guaranteed to be satisfied within a single round.}

For inter-shard communication between shards $S_1$ and $S_2$, we used broadcast-based protocol \cite{hellings2022fault} where a set $A_1 \subseteq S_1$ of $f_1+1$ nodes in $S_1$ and a set $A_2 \subseteq S_2$ of  $f_2 + 1$ nodes in $S_2$ are chosen (where $f_i$ is the number of faulty nodes in shard $S_i$). Each node in $A_1$ is instructed to broadcast the message to all nodes in $A_2$. Thus, at least one non-faulty node in $S_1$ will send the correct message value to a non-faulty node in $S_2$.


We assume that the nodes in a shard are close to each other and connected in a local area network. However, the distance between shards may vary.
We model the interconnection network between shards as a weighted complete graph (clique graph)
of shards $G_s$.
We measure the distance between shards as the number of rounds that are needed 
until a message is delivered over the network,
where a round is the time to reach consensus within a shard.
We consider two communication models:
\begin{itemize}
\item
{\em Uniform communication model}: Any two shards are a unit distance away,
in the sense that any shard can send or receive information within one round. In other words, the shards form a clique where each edge has a weight of $1$.

\item
{\em Non-uniform communication model:} the distance between any two shards ranges from $1$ to $D$, where $D$ is the diameter of the clique.
Hence, the edge weights vary from $1$ to $D$. We can simply say $D$ is the upper bound on the time needed to deliver a message from one shard to another. The unit of communication time is round, which is the time needed to reach a consensus within a shard.
\end{itemize}
Note that the uniform communication model is a special case of the
the non-uniform model where the distance of every edge is 1.

Each shard maintains its own local ledger (local blockchain) based on the subtransactions it receives. \ra{Moreover, in our algorithms, we consider a simple block structure where each block contains only one transaction. However, our algorithms can be extended to accommodate multiple transactions per block.}
We denote by $f_i$ is the number of Byzantine nodes in shard $S_i$.
We assume that each shard runs the PBFT \cite{PBFT} consensus algorithm to ensure agreement on the state of the local ledger.
To achieve Byzantine fault tolerance, the number of nodes in each shard must satisfy $n_i > 3 f_i$. 
Whenever it is required, it is possible to combine and serialize the local chains to form a single global blockchain~\cite{adhikari2023lockless}.


Consider a set of shared accounts $\mathcal{O}$
(we also refer to the accounts as {\em objects}). 
As in previous studies \cite{Byshard, adhikari2023lockless}, 
we assume that each shard is responsible for a specific subset of the shared objects (accounts).
Namely, $\mathcal{O}$ is divided into disjoint subsets $\mathcal{O}_1, \ldots, \mathcal{O}_s$, where $\mathcal{O}_i$ represents the set of objects managed by shard $S_i$.
Each shard $S_i$ maintains a local blockchain of subtransactions that access objects in the respective $\mathcal{O}_i$. 


\paragraph{\bf Transactions and Subtransactions}

Consider a transaction $T_i$. 
A transaction is injected into one of the nodes of the system, say node $v_{T_i}$.
The {\em home shard} of $T_i$ is the shard that contains $v_{T_i}$.
Each shard that receives newly generated transactions acts as a home shard for those transactions. 
A home shard maintains a {\em pending transactions queue}, which contains any newly generated transactions that were injected into it.
The home shard is responsible for handling all pending transactions in its queue.

We define a transaction $T_i$ as a collection of subtransactions $T_{i,a_1},\ldots,T_{i,a_j}$. 
Each subtransaction $T_{i,a_l}$ accesses objects 
only in $\mathcal{O}_{a_l}$,
and is associated with shard $S_{a_l}$.
Thus, subtransaction $T_{i,a_l}$ has a respective {\em destination shard} $S_{a_l}$.
The home shard of $T_i$ will send subtransaction $T_{i,a_l}$ to shard $S_{a_l}$ for processing,
where $T_{i,a_l}$ will be appended into the local blockchain of $S_{a_l}$.
%
The subtransactions within a transaction $T_i$ are independent 
(i.e. they do not conflict, as explained below) 
and can be processed concurrently.
Similar to previous work \cite{Byshard}, each subtransaction $T_{i,a_l}$
has two parts: (i) a condition check, where it checks 
whether a condition of the objects in $\mathcal{O}_{a_l}$
is satisfied, and (ii) the main action, where it updates the 
values of the objects in $\mathcal{O}_{a_l}$.
%
\begin{example}
Consider a transaction $T_1$ consisting of read-write operations on the accounts with several conditions.
$T_1$ = ``Transfer 1000 from Rex's account to Alice's account, if Rex has 5000 and Alice has 200 and Bob has 400''. The home shard of $T_1$ splits this transaction into three subtransactions $T_{1,r}, T_{1,a}, T_{1,b}$, where the destination shards $S_{r}$, $S_{a}$, and $S_{b}$ handle the respective accounts of Rex, Alice, and Bob:
\end{example}
\begin{itemize}[itemindent=0.7cm]
\item[$T_{1,r}$] - condition: ``Check Rex has 5000''
\item[]  - action: ``Remove 1000 from Rex account''
\item[$T_{1,a}$] - condition: ``Check Alice has 200''
\item[] - action: ``Add 1000 to Alice account''
\item[$T_{1,b}$] - condition: ``Check Bob has 400''
\end{itemize}
The home shard of $T_1$
sends the subtransactions to their respective destination shards. If the conditions are satisfied (for example in $T_{1,r}$ if Rex has 5000) and the transaction is valid (for example in $T_{1,r}$ Rex has indeed 1000 in the account to be removed) then the destination shards are ready to commit the subtransactions in the local blockchains, which imply that the whole of transaction $T_1$ implicitly commits as well. Otherwise, if any of the conditions in the subtransactions are not satisfied or the subtransactions are invalid, then 
all the subtransactions must abort (i.e. they are not added in the local blockchains), which results in $T_1$ aborting as well.

Transactions $T_i$ and $T_j$ are said to {\em conflict} if they access some object $O_l \in \mathcal{O}$ and at least one of these transactions 
writes (updates) the value of object $O_l$. 
Transactions that conflict should be processed in a sequential manner
to guarantee atomic object updates.
In such a case, their respective subtransactions 
should serialize in the exact same order 
in every involved shard to ensure atomicity of transaction execution.
In our algorithms, we construct the commit schedule 
with the help of a conflict graph of the transactions.
A transaction conflict graph $G$ for a set of transactions ${\T}$ is an unweighted graph where each transaction $T_i \in {\T}$ corresponds to a node of $G$ and an edge between any two transactions $T_i$ and $T_j$ that conflict.
In our algorithms, we will perform a vertex coloring of graph $G$ 
to produce a conflict-free schedule.


\paragraph{\bf Adversarial Model}
\label{subsec:adversarial-model}
We examine an adversarial model where transactions enter into the system continuously. 
The adversary generates and injects transactions into the system
with {\em injection rate} $\rho$, where $0 < \rho \leq 1$, and {\em burstiness} $b > 0$.
Each injected transaction adds {\em congestion of one unit} to each shard it accesses,
where the congestion of a shard measures the number of transactions that access objects in the shard.
The adversary is restricted such that the congestion on each shard 
within a contiguous time interval of duration $t > 0$
is limited to at most $\rho t + b$ transactions per shard.
While $\rho$ models a bound on the injection rate,
the burstiness parameter $b$ expresses the maximum number of transactions 
that the adversary can arbitrarily generate in any time~interval.

\paragraph{\bf Performance Metrics}
A scheduling algorithm 
responsible for determining the order in which transactions are processed within the sharded blockchain.
The execution of our proposed algorithm is synchronous, where the execution timeline is partitioned into time steps referred to as {\em rounds}. 
We assume that the duration of a round is enough to allow 
the execution of the PBFT consensus algorithm in each shard.
A round is also the time to send a message between shards in a unit distance.
The primary goal of a scheduling algorithm is to efficiently and fairly process all generated transactions while minimizing latency, ensuring the stability and performance of the sharded blockchain~system. 

A scheduler is considered stable with respect to the adversary if the number of pending transactions remains bounded throughout any execution (bounded with respect to the system parameters). In this context, stability implies that the scheduler can handle incoming transactions without an unbounded accumulation of pending transactions.
At any round, exactly one subtransaction can be processed in each shard.
Thus, the maximum congestion for a shard in $r$ rounds should not exceed $r$.
That is, if the injection rate of a scheduler exceeds 1, no scheduler can achieve stability. Therefore, we focus on adversaries where $0 < \rho \leq 1$.

The delay of a transaction $T$ refers to the number of rounds between its generation and moment of {\em commit},
where all of its subtransactions have been appended to the respective local blockchains. The latency of a scheduler in a particular execution is defined as the maximum delay among all transactions generated in that execution. In our algorithms, we bound the latency with respect to the parameters of the system.

\section{Upper Bound on Stable Generation Rate of Transactions}
\label{sec:lower-bound}

\dk{If the transaction generation rate is sufficiently high, the system can become unstable. This section gives an upper bound on the maximum transaction generation rate $\rho$ under which any blockchain sharding system could be stable. }
The following result is adapted from \cite{busch2023stable},
which originally considered transactions in software transactional memory systems.
Here, we adapted the result in the context of blockchain sharding.

\begin{theorem}[\dk{Stability Upper} Bound]
No transaction scheduler in any sharded blockchain system can be stable 
if the (worst-case adversarial) transaction generation rate $\rho$ satisfies $\rho > max\{ \frac{2}{k+1},  \frac{2}{ \left\lfloor\sqrt{2s}\right\rfloor}\}$ and burstiness $b>0$, 
where each transaction accesses at most $k$ out of $s$ shards.
\end{theorem}

\begin{proof}
Let us consider a \dk{sharded blockchain system}
with $s$ shards.
We consider the case where each shard holds one account, and each transaction accesses at most $k$ shards out of the $s$ shards. 
We analyze two cases. 

{\bf Case 1:} $\frac{k(k+1)}{2} \leq s$: The generated transactions use only shards $S_1, S_2, \dots, S_r$, where $r= \frac{k(k+1)}{2}$. We form a set of transactions, $T_1, T_2, \dots, T_{k+1}$, where each transaction $T_i$ accesses a subset of $k$ shards 
so that two transactions conflict with each other.
Specifically,
every pair of transactions maps to a unique shard used by both transactions. 
Note that this shard uniqueness can be guaranteed by the fact that the number of all transactions' pairs, $\frac{(k+1)k}{2}$, is not bigger than the number of shards, $s$.
Since the transactions conflict mutually, 
only one transaction can be committed in any round. 
Thus, the transactions require $k + 1$ rounds for all to be committed. 
The given group of $k+1$ transactions together contribute $2$ to the congestion of each used shard.
Hence, to ensure stability, 
the inequality  $\rho\cdot (k + 1) \leq 2$ must hold,
which implies $\rho \leq \frac{2}{k+1}$.

{\bf Case 2:} $\frac{k(k+1)}{2} > s$, let $p$ be the greatest positive integer satisfying $\frac{p(p+1)}{2} \leq s$. We can adapt a similar argument as in the case $\frac{k(k+1)}{2} \leq s$. Specifically, we consider a set of transactions $T_1, T_2, \dots, T_{p+1}$ such that 
every pair of transactions maps to a unique shard. The inequality $\rho \leq \frac{2}{p+1}$ can still be obtained through the same reasoning as above. Furthermore, $\frac{p(p+1)}{2} \leq s$ implies $p+1 = \lfloor \frac{1}{2}(1+ \sqrt{1+8s})\rfloor$. By making some algebraic deductions, we estimate $\frac{2}{p+1}=\frac{2}{\lfloor \frac{1}{2}(1+\sqrt{1+8s})\rfloor} \leq \frac{2}{\lfloor \sqrt{2s} \rfloor}$.



Hence, if $\rho$ exceeds $\frac{2}{\lfloor \sqrt{2s}\rfloor}$, it also exceeds $\frac{2}{p+1}$. Consequently, an adversary with $\rho$ 
exceeding this threshold 
(and with $b\ge 1$)
can generate at least one transaction per round for each round $r$, and it can generate two transactions at some round after $r$, causing the pending transaction queue at some shard to grow unbounded.
\end{proof}

\section{Scheduler for Uniform Model}
\label{sec:basic-distributed-scheduler}

We first consider the uniform model, where 
 any shard can send a message 
to any other shard within one round.
Algorithm~\ref{alg:basic-distributed-scheduler}
is the basic distributed algorithm to schedule the transactions in the uniform model. 
The algorithm does not need to know the parameters $\rho$ and $b$
of the adversary.
All the shards are synchronized.
The basic idea is that the algorithm divides time into epochs,
and in each epoch, it first constructs the conflict graph $G$ of the generated transaction and then schedules the transactions in a conflict-free manner by using a coloring of $G$.

In each epoch, one of the shards acts as a {\em leader shard} $S_{\text{ldr}}$, 
which receives information about transactions from all the home shards.
The leader shard $S_{\text{ldr}}$ colors the transactions and then sends the coloring information back to the home shards. 
The leader shard is updated in each epoch using the modulo division on epoch number and the total number of shards, $S_{\text{ldr}} \leftarrow S_{(epoch \mod s) +1}$. 
All shards know the leader shard for the current epoch. 
The update of the leader shard in each epoch ensures fair load balancing to the shards.

\begin{figure}[t]
\centering
\includegraphics[width=0.48\textwidth]{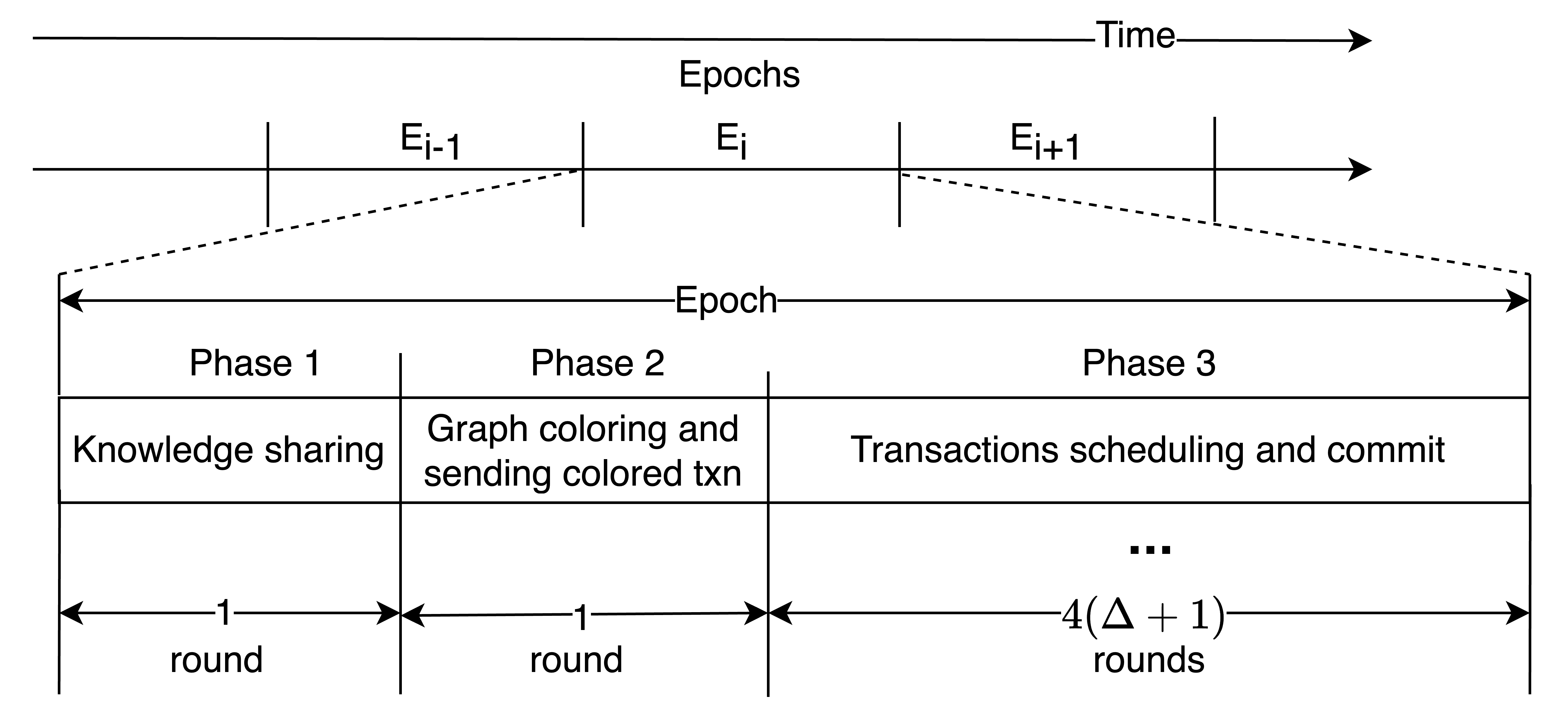}
\caption{Representation of time slots division in Algorithm~\ref{alg:basic-distributed-scheduler}}
\label{fig:time-slot-division}
\end{figure}

Each epoch consists of several rounds,
where the specific duration varies according to the load of pending transactions.
In each epoch, the home shards process all the pending transactions that appear in their pending queues at the beginning of the epoch.
Processing of these transactions occurs in three phases during the epoch
(see Figure \ref{fig:time-slot-division}).
In the first phase, the home shards communicate with the leader shard ($S_{\text{ldr}}$)
to discover conflicts
among the transactions.
In the second phase, 
the leader shard $S_{\text{ldr}}$ builds a conflict graph $G$ of the transactions
and uses a vertex graph coloring algorithm to color $G$,
then send the transaction coloring information back to the respective home shards.
Finally,
in the third phase,
each transaction $T$
is assigned a specific round based on its color,
and its subtransactions are sent to the destination shards
where all the subtransactions of $T$ commit concurrently at the
designated round in a conflict-free manner. 

Let $E_1, E_2, E_3, \ldots$ denote the sequence of epochs.
The number of rounds required to process transactions in a particular epoch $E_i$ depends on the maximum degree $\Delta$ of 
the conflict graph $G$ on that epoch.
In particular, Phases 1 and 2 consist of one round, and Phase 3 consists of $4(\Delta+1)$ rounds. Here, we need to multiply $\Delta+1$ by $4$ because the home shards and destination shards need to communicate with each other (back and forth) to check the validity of the transactions and subtransactions and for consistent and atomic commitment in each destination shard. 

\begin{algorithm}[t]
\small
\caption{Basic Distributed Scheduler (BDS)}
\label{alg:basic-distributed-scheduler}
$S_{\text{ldr}} \gets$ leader shard, updates in each epoch\;
\BlankLine

 \SetKwBlock{DoInRound}{\normalfont {{\bf foreach} {\em epoch and appropriate round} {\bf do}}}{}
    \DoInRound{
       \tcp{Phase 1: Knowledge Sharing}
         Each home shard picks all pending txn (transactions) from its queue and sends them to $S_{\text{ldr}}$ shard for coloring\;

        \BlankLine
        \tcp{Phase 2: Graph Coloring in leader shard $S_{\text{ldr}}$}
       $S_{\text{ldr}}$ colors received txn with at most $\Delta+1$ colors\;
       $S_{\text{ldr}}$ sends colored txn to respective home shards\;
     Each home shard inserts received colored txn in pending queue\;
        \BlankLine
        \tcp{Phase 3: Schedule and commit}
        \tcc{color $col$ is processed at appropriate round(s)}
        \For{color $col \leftarrow 1$ \KwTo $\Delta+1$}{
            \tcc{Round 1: Home Shards}
            Home shard picks each txn $T_i$ of color $col$ from pending queue 
            and splits it into subtransactions ($T_{i,j}$) which are sent to destination shards for voting\;
            
            \BlankLine
            \tcc{Round 2: Destination shards}
            Send {\em commit vote} to home shard if $T_{i,j}$ is valid and condition is satisfied; otherwise, send {\em abort vote}\;
            
            \BlankLine
            \tcc{Round 3: Home Shards}
            Send {\em confirm commit} to destination shards if all ``commit votes'' are received; otherwise, send {\em confirm abort}\;
            
            \BlankLine
            \tcc{Round 4: Destination Shards}
            If destination shard receives ``confirm commit'', it commits $T_{i,j}$ by appeding it in local blockchain; 
            Otherwise, if it receives ``confirmed abort'', it aborts $T_{i,j}$\;
        }
        
    }  
\end{algorithm}

We continue to describe the phases in more detail.

{\bf Phase 1:} (Knowledge sharing) 
In this phase, the home shards send all transaction information from their pending queues to the leader shard $S_{\text{ldr}}$ in parallel using one round.

{\bf Phase 2:} (Graph coloring)
In this phase, the leader shard $S_{\text{ldr}}$ creates a global conflict graph $G$ using the received transaction information from each home shard.
Let $\Delta$ be the degree of $G$.
Then, 
$S_{\text{ldr}}$ shard uses a vertex coloring algorithm to color the conflict 
graph with at most $\Delta+1$-colors.
(We assume the system has enough resources to color the graph in the beginning of the round.) 
Then $S_{\text{ldr}}$ sends the colored transaction information to the respective home shards for scheduling based on the colors.

{\bf Phase 3:} (Transactions scheduling and committing)
Transactions that obtain the same color are conflict-free and can be scheduled to commit at the same round. To ensure the atomicity of transactions at each destination shard, the home shard and destination shards need to communicate with each other, which takes four rounds per color. Thus, all transactions can be committed within a maximum of $4(\Delta+1)$ rounds. 

The home shard of a transaction $T_i$ with color $z$ processes the transaction at round \ra{$4z$} of the third phase.
Each color needs four rounds. 
In the first round, each home shard splits $T_i$ into subtransactions
and sends them to the destination shards to check the condition and constraints.
In the second round, if a subtransaction $T_{i,j}$ is valid and the condition is satisfied,
then the destination shard sends {\em commit vote}, or otherwise {\em abort vote}, for $T_{i,j}$ to its home shard.
In the third round, if the home shard receives all the commit votes for $T_i$, then it sends {\em confirmed commit} to the respective destination shards; otherwise if any abort vote is received, it sends {\em confirmed abort}. In the fourth round, the destination shard of each subtransaction $T_{i,j}$ either commits $T_{i,j}$ and adds to the local blockchain or aborts $T_{i,j}$ according to the message from the home shard.

\subsection{Analysis for Algorithm \ref{alg:basic-distributed-scheduler}}
For the analysis,
consider the case where
each transaction accesses at most $k \geq 1$ out of $s \geq 1$ available shards.

\begin{lemma}
\label{lemma:distributed-epoch-without-range}
In Algorithm \ref{alg:basic-distributed-scheduler},
for generation rate $\rho \leq max\{ \frac{1}{18k}, \frac{1}{18\lceil \sqrt{s} \rceil} \}$ and burstiness $b \geq 1$,
every epoch satisfies:
(i) the maximum epoch length is $\tau := 18b\cdot min \{k, \lceil \sqrt{s}\rceil\}$,
and (ii) at most $2bs$ new transactions are generated during the epoch. 
\end{lemma}

\begin{proof}
    We start by noting that if an epoch $E_i$ 
    has length at most $\tau$,
    and the adversary generates transactions with a rate $\rho \leq max\{ \frac{1}{18k}, \frac{1}{18\lceil \sqrt{s} \rceil} \}$ and burstiness $b \ge 1$, 
    the maximum congestion added to a shard during the epoch is:
    \begin{equation}
    \label{eqn:transaction-generation-rate}
    \rho \cdot  \text{\em Epoch length} +b \leq \rho \cdot \tau +b \leq b+b=2b.
    \end{equation}
    Moreover, since each transaction accesses at least one shard and there are $s$ shards,
    the total number of transactions produced within epoch $E_i$ is at most $2bs$.
    Similarly, $2bs$ is the maximum total congestion in $E_i$.
    Hence, property (i) implies property (ii). Thus, it suffices to prove property (i).

    We continue to prove property (i) by induction on the number of epochs.
    For the base case, in epoch $E_1$, there are no transactions at the beginning,
    and the epoch length is just two rounds. Thus property (i) 
    holds since $2 \leq \tau$.
    
    Assume the property (i) holds for every epoch up to $E_j$.
    For the inductive step, consider epoch $E_{j+1}$. 
    By design of Algorithm \ref{alg:basic-distributed-scheduler},
    every transaction which is pending in the beginning of $E_{j+1}$ was generated during epoch $E_j$;
    we refer to these as ``old transactions''.
    During $E_{j+1}$, only the old transactions are scheduled.
    We examine two cases.
   
   {\bf (Case 1) $k \leq \lceil \sqrt{s} \rceil$:}
   From Equation \ref{eqn:transaction-generation-rate}, 
   the congestion of a shard is at most $2b$ old transactions. 
   Since each old transaction accesses at most $k$ shards, 
   each old transaction conflicts with at most $(2b-1)k$ other old transactions. Consequently, the highest degree $\Delta$ in the conflict graph $G$ of the old transactions is $\Delta \leq (2b-1)k$. Thus, in the second phase of $E_{j+1}$,
   a (greedy) vertex coloring algorithm on $G$ assigns 
   at most $\Delta + 1 \leq (2b-1)k + 1$ colors. 
   Since $k \geq 1$, the total length of epoch $E_{j+1}$ is at most
   $$2 + 4(\Delta+1) \leq 2 + 4((2b-1)k + 1) < 18 b k = \tau.$$
   Therefore, property (i) holds.

    {\bf (Case 2) $k> \lceil \sqrt{s}\rceil$:}
     we classify old transactions into two groups, the ``heavy'' transactions which access more than $\lceil \sqrt{s}\rceil$ shards, and the ``light'' transactions which access at most $\lceil \sqrt{s}\rceil$ shards.
     The maximum number of heavy old transactions can be $2b\lceil \sqrt{s}\rceil$. If there were more, the total congestion of old transactions would be strictly greater than $2b\lceil \sqrt{s}\rceil \cdot \sqrt{s} \geq 2bs$, which is not possible.
     A coloring algorithm for conflict graph $G$ can assign each of the heavy transactions a unique color, which requires at most 
     \ra{$\zeta_1 =2b\lceil \sqrt{s}\rceil$ colors}.

     The remaining transactions of $G$ are light. 
     Let $G'$ be the subgraph of $G$ with the light transactions.
     Each light transaction conflicts with at most 
      $(2b-1)\lceil \sqrt{s}\rceil$
     other light transactions.
     Hence, 
     the degree of $G'$ is at most $(2b-1)\lceil \sqrt{s}\rceil$.
     Thus, $G'$ can be colored with at most $\zeta_2 = (2b-1)\lceil \sqrt{s}\rceil + 1$ colors.
     
     
     Consequently, $G$ can be colored with at most $\zeta = \zeta_1 + \zeta_2 = 2b\lceil \sqrt{s}\rceil + (2b-1)\lceil \sqrt{s}\rceil + 1$ colors.  
    Since $s \geq 1$, the length of $E_{j+1}$ is at most
    $$2  + 4 \zeta 
    = 6 + 4(4b - 1)\lceil \sqrt{s}\rceil \leq 18b \lceil \sqrt{s}\rceil = \tau\ \ .$$
    Therefore, property (i) holds.
\end{proof}

\begin{theorem}[BDS stability]
    In Algorithm \ref{alg:basic-distributed-scheduler}, for generation rate 
 $\rho \leq max\{ \frac{1}{18k}, \frac{1}{18\lceil \sqrt{s} \rceil} \}$ and burstiness $b \geq 1$, the number of pending transactions at any given round is at most $4bs$, and the transaction latency is at most $36b\cdot min \{k, \lceil \sqrt{s}\rceil\}$.
\end{theorem}
\begin{proof}
    To estimate the number of pending transactions during a round, consider a round within an epoch $E_i$. 
    From the proof of Lemma \ref{lemma:distributed-epoch-without-range}, the maximum number of old transactions during any round of $E_i$ is $2bs$. Within $E_i$, there can be at most $2bs$ newly generated transactions. Therefore, the upper bound on pending transactions during a round is $4bs$.

    For estimating transaction latency, we rely on the fact that a transaction generated in an epoch will be processed by the end of the next epoch. 
    Consequently, the transaction latency is bounded by twice the duration of the maximum epoch length $\tau$.
    From Lemma \ref{lemma:distributed-epoch-without-range},
    this results in a latency of at most 
    $$2 \tau = 2 \cdot 18b\cdot min \{k, \lceil \sqrt{s}\rceil\} = 36b\cdot min \{k, \lceil \sqrt{s}\rceil\}.$$
\end{proof}

\section{Fully Distributed Scheduler (DS)}
\label{sec:fully-distributed}

The scheduling algorithm we presented earlier uses a central authority in one shard with knowledge about all current transactions and the maximum degree of the transaction graph. Here, we discuss a \dk{fully} distributed approach that allows the transaction schedule to be computed in a decentralized manner without requiring a central authority \dk{in any shard}.

\subsection{Cluster Decomposition (Shard Clustering)}
\label{shard-cluster-decomposition}
This algorithm considers a non-uniform communication model described in Section \ref{preliminaries}. 
Consider the (complete) weighted shards graph $G_s$ made by the $s$ shards,
where the weights of edges between shards represent the distances between them.
We assume that $G_s$ is known to all the shards.
Let $D$ be the diameter of $G_s$ ($D$ is the maximum distance between two shards).
The $q$-neighborhood of shard $S_i$ is defined as the set of shards within a distance of at most $q$ from $S_i$, and the 0-neighborhood is simply $S_i$ itself.

The scheduling algorithm uses a hierarchical decomposition of $G_s$
which is calculated before the algorithm starts and is known to all the shards.
The graph decomposition is based on the clustering techniques in \cite{gupta2006oblivious} and which were later used in \cite{sharma2014distributed,busch2022dynamic}.
The hierarchy consists of $H_1 = \lceil \log D \rceil +1$ layers (logarithms are in base 2), where a layer is a set of clusters, and a cluster is a set of shards. 
Layer $l$, where $0\leq l < H_1$, is a sparse cover of $G_s$
such that: 
(i) each cluster of layer $l$ has (strong) diameter at most $O(2^l\log s)$;
(ii) each shard participates in no more than $O(\log s)$ clusters at layer $l$;
(iii) for every shard $S_i$ there exists a cluster at layer $l$ such that the entire $(2^l-1)$- neighborhood of $S_i$ is contained within that cluster. 
The clusters have a {\em strong diameter}, 
which measures distances within a cluster on the induced subgraph of $G_s$ induced by the nodes of the cluster.

For each layer $l$,
the sparse cover construction in \cite{gupta2006oblivious}
is actually obtained as a collection of $H_2 = O(\log s)$ partitions of $G_s$. 
These $H_2$ partitions are ordered as sub-layers of layer $l$ labeled from $0$ to $H_2-1$.
A shard participates in all $H_2$ sub-layers but potentially belongs to a different cluster at each sub-layer.
At least one of the clusters in one of these $H_2$ sub-layers at layer $l$ contains the whole $2^l-1$ neighborhood of $S_i$.

Within each cluster at layer $l$, a leader shard $S_{\text{ldr}}$ is specifically designated such that the leader’s $(2^l-1)$-neighborhood is in that cluster.
If we cannot find a leader (because no shard has its $(2^l-1)$-neighborhood) then no leader will be elected in the cluster.  No shard would have picked such a cluster as its home. Thus, that cluster will not be used at all.
Each transaction $T$ has a {\em home cluster} 
which is defined as follows: let $S_i$ be the home shard of $T$,
and $x$ the maximum distance from $S_i$ 
to the shards that will be accessed by $T$;
the home cluster of $T$ is the lowest-layer (and lowest sub-layer) cluster 
in the hierarchy that contains the whole $x$-neighborhood of $S_i$.
The designated leader shard $S_{\text{ldr}}$ of the home cluster will handle all the transactions that have their home shard in that cluster (i.e., transactions will move from the home shard to the leader shard $S_{\text{ldr}}$ for processing). 

\subsection{Fully Distributed Scheduling Algorithm}

Our scheduling algorithm runs in epochs. Each layer $i$ has a fixed epoch length $E_i = c2^{i}\log s$ rounds, for some constant $c$,
where $0 \leq i < H_1$.
Epochs repeat perpetually,
so that the next epoch starts right after the previous ends.
Note that the epochs of different layers are multiples
of each other,
that is, $E_i=2\cdot E_{i-1}$, for $i > 0$,
and their begin times are aligned so that the 
start of $E_{i-1}$ coincides with the start of $E_{i}$.
Notice all epoch lengths are multiples of the length $E_0 = c \log s$.

Similar to epochs,
we also define {\em rescheduling periods} $P_k = 2^k \cdot E_0$,
where $k \geq 0$.
Rescheduling periods are multiples of epoch $E_0$,
they repeat perpetually, 
and the start times are aligned.
Rescheduling periods are also aligned with the epochs.
For a cluster at layer $i$, the lowest rescheduling period 
is $P_{i} = 2^i E_0 = E_i$,
that is, the epoch length of layer $i$ is the same as the length of the lowest rescheduling period for that layer.

The reason for having epochs and rescheduling periods is as follows.
Under normal conditions, in a cluster $C$ at layer $i$, at each epoch, the designated leader shard $S_{\text{ldr}}$ colors and schedules transactions received from the home shards in $C$ which it then sends to the destination shards
for confirmation and commit.
However, some of the transactions may not commit right away at the destination shards
due to conflicts and may accumulate at the destination shard queues.
To avoid accumulating delays,
stale transactions are given additional chances to be rescheduled. 
Such rescheduling chances are given when 
the rescheduling periods align with the epochs.
When the end of the current epoch $E_i$ aligns with one of the rescheduling periods $P_k$, where $k > i$, then $S_{\text{ldr}}$ colors not only newly received transactions but also those that have already been scheduled but not yet confirmed (or not committed). 
This approach expedites the commitment of multiple non-conflicting transactions in parallel, reducing the overall transaction waiting time.

As discussed earlier, 
each cluster $C$ belongs at some layer $i$ and sublayer $j$,
where $0 \leq i < H_1$ and $0 \leq j < H_2$;
we simply say that $C$ is at {\em level} $(i,j)$.
Consider a transaction $T$ with home cluster $C$;
we also say that $T$ is at level $(i,j)$.
The leader shard $S_{\text{ldr}}$ of $C$ assigns an integer {\em color} to $T$,
during some occurrence of epoch $E_i$.
Let $t_{\text{end}}$ denote the end time of that epoch,
where $T$ got colored.
We introduce the concept of the {\em height} of $T$ represented by a tuple $(t_{\text{end}}, i, j, \text{color})$.
The heights of transactions are ordered lexicographically and implement a 
priority scheme. The priority for committing transactions is determined based on this order, giving precedence to transactions generated in lower cluster layers (with smaller epoch lengths). This prioritization strategy aims to leverage locality to improve the transaction latency.

We divided our fully distributed scheduling algorithm (see Algorithm~\ref{alg:fully-distributed-scheduler}) into two sub-algorithms:  Transaction Scheduling (Algorithm \ref{alg:fully-distributed-scheduler-a}) and Transaction Confirming and Committing (Algorithm \ref{alg:fully-distributed-scheduler-b}). In the following, we continue to describe each algorithm.


\renewcommand\theContinuedFloat{\alph{ContinuedFloat}} 
\begin{algorithm}[t]

\small
\caption{Fully Distributed Scheduler (FDS)}
\label{alg:fully-distributed-scheduler}
\BlankLine
 \SetKwBlock{DoParallel}{In parallel do}{}
\DoParallel{
         Run Algorithm \ref{alg:fully-distributed-scheduler-a} \tcp{Transactions Scheduling}
        Run Algorithm \ref{alg:fully-distributed-scheduler-b}\tcp{Transactions Committing }
    }
    
\end{algorithm}



    
    

\subsubsection{Transactions Scheduling (Algorithm \ref{alg:fully-distributed-scheduler-a})}

Every generated transaction $T$ has a {\em home cluster} at some respective level $(i,j)$.
Algorithm~\ref{alg:fully-distributed-scheduler-a} runs in epochs whose length depends on the level.
Each transaction is processed at the respective epoch for their level.
Each epoch consists of three phases, which we describe below.

{\bf Phase 1:} In this phase, at the beginning of each epoch, a home shard sends transactions to the cluster leader $S_{\text{ldr}}$ of the respective home cluster. Note that a home shard may have transactions at various levels.
Those transactions will be processed at the respective level.

{\bf Phase 2:} In this phase, the cluster leader $S_{\text{ldr}}$ colors the transactions received from home shards.
If the end time of the current epoch $E_i$ aligns with one of the rescheduling periods $P_k$, $k > i$, then $S_{\text{ldr}}$ colors all transactions ever received from the home shard, including those already scheduled but not yet committed (excluding the already committed). However, if the current epoch does not align with a rescheduling period, then $S_{\text{ldr}}$ only colors the newly received transactions, excluding already scheduled transactions.
For the purpose of rescheduling, 
the leader shard maintains a queue $sch_{\text{ldr}}$ which contains 
all uncommitted transactions it has processed.

After completing the coloring operation,
$S_{\text{ldr}}$ splits
colored transactions into subtransactions and then sends them to destination shards for scheduling. 
Additionally, $S_{\text{ldr}}$ informs the destination shards about whether their destination schedule queue $sch_{\text{qd}}$ needs updating or appending based on whether the received subtransaction results from rescheduling or regular scheduling. This ensures destination shards appropriately update their schedule queues.

{\bf Phase 3:} In this phase, each destination shard appends or updates the scheduled subtransactions in their queues according to the received information from cluster leader shard $S_{\text{ldr}}$. Additionally, each destination shard lexicographically sorts subtransactions based on their latest height $(t_{\text{end}}, i, j, color)$. Thus, transactions and subtransactions have the same relative order in all destination shards, ensuring consistent order. 
%

\begin{algorithm}[t]
\ContinuedFloat
\small
\caption{Transactions Scheduling}\label{a:2a}
\label{alg:fully-distributed-scheduler-a}


$sch_{\text{ldr}} \leftarrow$ scheduled transaction (txn) queue in each leader shard\;
$sch_{\text{qd}} \leftarrow$ scheduled subtransaction queue in each destination shard\;



For every new transaction $T$, the home shard assigns  the {\em home cluster} of $T$\;

\BlankLine
\SetKwBlock{DoParallel}{For all layers ($i,j$) in parallel do}{end}
    \DoParallel{
\ForEach {epoch $E_i$ of layer $i$ and appropriate round} {
        
        
        \tcp{Phase 1: txn sending}
        Home shard of $T$ sends transaction $T$ to the leader shard $S_{\text{ldr}}$ of the {\em home cluster} of $T$\;

        \BlankLine
        \tcp{Phase 2: Txn Coloring in $S_{\text{ldr}}$ }
        \If {end time $t_{\text{end}}$ of $E_i$ coincides with end time of one of the rescheduling periods $P_k$, $k > i$}{
             Leader shard $S_{\text{ldr}}$ colors all txn in its queue $sch_{\text{ldr}}$ including the transactions it receives during current epoch $E_i$ from the home shards\;
             $S_{\text{ldr}}$ updates $sch_{\text{ldr}}$ with all the colored transactions\; 
        }
        \Else {
             $S_{\text{ldr}}$ colors all txn it received from home shards during current epoch $E_i$ 
             and appends them to $sch_{\text{ldr}}$\;
        } 
            $S_{\text{ldr}}$ splits colored txn into subtransactions and sends them to the respective destination shards for scheduling\;
        
        \BlankLine
        \tcp{Phase 3: Scheduling txn in destination shards}
             Destination shards update or append $sch_{\text{qd}}$ with received subtransactions according to message from $S_{\text{ldr}}$\;
            
         Order subtransactions in $sch_{\text{qd}}$ lexicographically using height $(t_{\text{end}}, i, j, color)$\;

 
    }
}
    
\end{algorithm}

\subsubsection{Transactions Confirming and Committing (Algorithm \ref{alg:fully-distributed-scheduler-b})}
Algorithm \ref{alg:fully-distributed-scheduler-b} is responsible for committing subtransactions in the destination shard queues, which were scheduled by Algorithm \ref{alg:fully-distributed-scheduler-a}.
Algorithm \ref{alg:fully-distributed-scheduler-b} runs at the destination shards independently and in parallel with Algorithm \ref{alg:fully-distributed-scheduler-a}.
The committing time of transactions depends on which level $(i,j)$ that transaction belongs to.
If the diameter of the home cluster of transaction $T_r$ is $d_r$, then it will take at most $2d_r+1$ rounds to commit because the cluster leader shard and destination shard need to communicate back and forth for confirmation of the transaction $T_r$, where step $1$ and $2$ each takes $d_r$-rounds and step $3$ takes $1$ round. We describe each step of Algorithm \ref{alg:fully-distributed-scheduler-b} as follows:

{\bf Step 1:} In this step, each destination shard picks one subtransaction from the head of the scheduled queue $sch_{\text{qd}}$ and checks the condition and validity. If everything checks out, the shard casts a commit vote and sends it to cluster leader shard $S_{\text{ldr}}$; otherwise, it sends an abort vote to $S_{\text{ldr}}$ for that subtransaction.

{\bf Step 2:} In this step, cluster leader shard $S_{\text{ldr}}$ collects all subtransaction votes for a specific transaction and sends a commit confirmation to the destination shard upon receiving all commit votes. Conversely, if the cluster leader shard receives any abort vote for a particular transaction, it sends an abort confirmation message. After sending a confirmed commit or abort message, $S_{\text{ldr}}$ removes that transaction from its own queue $sch_{\text{ldr}}$.

{\bf Step 3:} In this step, each destination shard makes a final commitment or abort based on the message received from the cluster leader shard and removes that subtransaction from its schedule queue ($sch_{\text{qd}}$).

\begin{algorithm}[t]
 \ContinuedFloat
\small
\caption{Transactions Confirming and Committing }\label{b:2b}
\label{alg:fully-distributed-scheduler-b}
$sch_{\text{qd}} \leftarrow$ schedule queues in destination shards\;
 \BlankLine 
 \SetKwBlock{DoInRound}{\normalfont {{\bf foreach} {\em shard } {\bf do}}}{end}
    \DoInRound{
      
        \tcp{Step 1: Condition check at destination shards:}
         $T_{i,j}\leftarrow$ pick one subtransaction from the head of $sch_{\text{qd}}$\;

         Send {\em commit vote} to  $S_{\text{ldr}}$ shard if $T_{i,j}$ is valid and condition is satisfied; otherwise, send {\em abort vote} for $T_{i,j}$ to $S_{\text{ldr}}$\;

        \BlankLine
        \tcp{Step 2: Collect votes at leader $S_{\text{ldr}}$ shards:}
         For each txn $T_i$, $S_{\text{ldr}}$ collects the votes about the subtransactions of $T_i$ 
         received from destination shards at Step 1\;

        Send {\em confirmed commit} $T_{i,j}$ to destination shards if all ``commit votes'' are received for $T_i$ and then remove $T_i$ from $sch_{\text{ldr}}$\;

        Send {\em confirmed abort} $T_{i,j}$ to destination shards if any ``abort vote'' is received for $T_i$ and then remove $T_i$ from $sch_{\text{ldr}}$\;

        \BlankLine
        \tcp{Step 3: Commit SubTxn at destination Shards:}
        Commit (or abort) $T_{i,j}$ and add to local blockchain (or abort) $T_{i,j}$ according to the message received from $S_{\text{ldr}}$ at Step 2\;

        If $T_{i,j}$ gets committed or aborted then remove $T_{i,j}$ from $sch_{\text{qd}}$
    }

\end{algorithm}

\subsection{Analysis for Algorithm \ref{alg:fully-distributed-scheduler}}

For analysis, consider a case where each transaction accesses at most $k \geq 1$ out of $s \geq 1$ shards and burstiness $b\geq 1$.
Let $C$ be a cluster at level $(i,j)$,
and let $d_i$ be the cluster diameter.
Let $\tau_i := 15bd_i\cdot min\{k, \sqrt{s}\}$.
Let $P_k$ be a rescheduling period with $\tau_i \leq P_k \leq 2\tau_i$.
Consider a sequence of such rescheduling periods
$I_1, I_2, \ldots$, each of length $P_k$.

\begin{lemma}
\label{lemma:transaction-schedule-alg3}
In Algorithm \ref{alg:fully-distributed-scheduler},
assuming that only transactions with home cluster $C$ are generated,
for 
$\rho \leq max\{ \frac{1}{30d_i k}, \frac{1}{30d_i\lceil \sqrt{s} \rceil} \}$
and $b\geq 1$,
in any period $I_z$, where $z \geq 1$:
(i) at most $2bs$ new transactions are generated with home cluster $C$, and
(ii) all these transactions 
will be committed or aborted by the end of $I_{z+1}$.
\end{lemma}

\begin{proof}
    During any of the periods $I_z$, since its length is at most $2\tau_i$, it follows that the maximum congestion added to any shard in C is:
    \begin{equation}
    \label{eqn:transaction-generation-rate-alg3}
    \rho \cdot P_k + b \leq 2 \rho \tau_i + b \leq b+b=2b.
    \end{equation}
    Moreover, since each transaction accesses at least one shard and there are at most $s$ shards in $C$, the total number of transactions produced in $I_z$ is at most $2bs$.
    Thus, the newly generated transactions with home cluster $C$ are also at most $2bs$. Hence, property (i) holds.

    We continue to prove property (ii).
    We prove it by induction on the number of periods.
    To simplify, let $I_0$ be a trivial first period (and not $I_1$), 
    such that no transactions were generated in $I_0$. 
    Then, there is nothing to schedule in $I_1$,
    and the basis case trivially holds.
    
    Suppose that property (ii) holds for all 
    periods from $I_0$ up to $I_{z-1}$. 
    We will prove property (ii) for $I_{z}$.

    By induction hypothesis,
    every transaction which is pending at the beginning of $I_{z+1}$ was generated during interval $I_z$; we refer to these as ``old transactions''.
    Here, we are considering only one cluster $C$ with one level. Thus, other levels will not affect the processing of old transactions.
    We examine two cases based on the magnitude of $k$ and $\lceil\sqrt{s}\rceil$.

     {\bf (Case 1) $k \leq \lceil \sqrt{s} \rceil$:}
    From Equation \ref{eqn:transaction-generation-rate-alg3}, the congestion of transaction to a shard is $2b$ old transactions. Since each transaction accesses at most $k$ shards, each old transaction conflicts with at most $(2b-1)k$ other old transactions. Consequently, the highest degree $\Delta$ in the conflict graph $G$ of the old transactions is $\Delta \leq (2b-1)k$. Thus, in Phase 2 of Algorithm \ref{alg:fully-distributed-scheduler-a} executed in the final epoch of period  $I_{z+1}$, a greedy vertex coloring algorithm on $G$ assigns at most $\Delta+1 \leq (2b-1)k+1$ colors. Since $k \geq 1$, the total length of the required time interval to schedule and commit/abort the transactions is calculated as follows. Scheduling Algorithm \ref{alg:fully-distributed-scheduler-a} consists of three phases where Phase 1 and Phase 2, each takes $d_i$ rounds, and Phase 3 takes $1$ round for transaction scheduling resulting $2d_i+1$ rounds. Similarly, Algorithm \ref{alg:fully-distributed-scheduler-b} takes $2d_i + 1$ rounds for each color to confirm and commit the transaction. 
    $$2d_i+1 + (2d_i+1)(\Delta+1) \leq 2d_i + 1+ (2d_i+1)((2b-1)k + 1) < 15bd_ik = \tau_i \ .$$
    
    Thus, property (i) holds.

     {\bf (Case 2) $k> \lceil \sqrt{s}\rceil$:}
     we classify old transactions into two groups, the ``heavy'' transactions, which access more than $\lceil \sqrt{s}\rceil$ shards, and ``light'' transactions, which access at most $\lceil \sqrt{s}\rceil$ shards.
       The maximum number of heavy old transactions can be $2b\lceil \sqrt{s}\rceil$. If there were more, the total congestion of old transactions would be strictly greater than $2b\lceil \sqrt{s}\rceil \cdot \sqrt{s} \geq 2bs$, which is not possible.
    A coloring algorithm for conflict graph $G$ can assign each of the heavy transactions a unique color, which requires at most $\zeta_1 =2b\lceil \sqrt{s}\rceil$ colors.

    The remaining transactions of $G$  are light. Suppose $G'$ be the subgraph of $G$  with the light transactions. Each light transaction conflicts with at most $(2b-1)\lceil \sqrt{s}\rceil$
     other light transactions.
     Hence, 
     the degree of $G'$ is at most $(2b-1)\lceil \sqrt{s}\rceil$.
     Thus, $G'$ can be colored with at most $\zeta_2 = (2b-1)\lceil \sqrt{s}\rceil + 1$ colors.

    Consequently, $G$ can be colored with at most $\zeta = \zeta_1+\zeta_2 = 2b\lceil \sqrt{s}\rceil + (2b-1)\lceil \sqrt{s}\rceil + 1$ colors. Since $s\geq 1$, the length of $I_{z+1}$ is at most:
     $$2d_i+1 + (2d_i+1) \zeta 
    = 4d_i +2+ (2d_i+1)(4b - 1)\lceil \sqrt{s}\rceil \leq 15bd_i \lceil \sqrt{s}\rceil = \tau_i\ \ .$$
   
    Since the process was started before the start of period $I_{z+1}$, and $I_{z+1}$ has length at least $\tau_i$, it follows that property (ii) holds
\end{proof}

We will now consider transactions from all levels.
Let $i'$ be the maximum layer accessed by any transaction
where the diameter is at most $d_{i'}$.
Let $\tau_i := 30 bd_{i} H_2 \cdot min\{k, \sqrt{s}\}$.
Consider now a sequence of such rescheduling periods
$I_1, I_2, \ldots$, each of length $P_k$,
where $\tau_{i'} \leq P_k \leq 2 \tau_{i'}$.

\begin{lemma}
\label{lemma:upper-bound-alg3}
In Algorithm \ref{alg:fully-distributed-scheduler},
for 
$\rho \leq max\{ \frac{1}{60d_{i'}H_2 k}, \frac{1}{60d_{i'}H_2\lceil \sqrt{s} \rceil} \}$
and $b\geq 1$,
in any period $I_z$, where $z \geq 1$,
for any destination shard:
(i) at most $2bs$ new transactions are generated (from all levels),
(ii) all these transactions 
will be committed or aborted by the end of $I_{z+1}$.
\end{lemma}

\begin{proof}
The total transaction congestion in each shard during a 
rescheduling period $I_z$ (which has length at most $2 \tau_{i'}$) 
is
$\rho \cdot 2 \tau_{i'} + b \leq b+b =2b$. 
Therefore, the total transaction weight for $s$ shards is at most $2bs$.
Hence, property (i) holds.

We continue with property (ii).
Consider a destination shard $S_q$.
From the proof of Lemma~\ref{lemma:transaction-schedule-alg3},
if we only had subtransactions from a layer $i$,
the schedule requires at most $\tau_i$ rounds
in $I_{z+1}$.
However, 
$S_q$ needs to process subtransactions 
from all layers $0,\ldots, i'$, and 
sublayers $0, \ldots, H_{2}-1$.
Those are executed with their assigned priorities.
since a layer $l$ cluster has a diameter $O(2^l\log s)$, as discussed in Section \ref{shard-cluster-decomposition}. Therefore, $d_l=O(2^l\log s)$. Let $d_l=c2^l\log s$, for some constant $c$. This implies $\sum_{i=0}^{i'}d_i\leq2d_{i'}$,
the total number of rounds needed to schedule all levels $(i,j)$ at $S_q$
is at most
\begin{align*}
\sum_{i=0}^{i'} \sum_{j=0}^{H_2-1} \tau_{i}
& =  \sum_{i=0}^{i'} \sum_{j=0}^{H_2-1} 15bd_i\cdot min\{k, \sqrt{s}\}\\
& \leq 30 H_2 b d_{i'} \cdot \min\{k, \sqrt{s}\} = \tau_{i'}\ .
\end{align*}
Since the length of $I_{z+1}$ is at least $\tau_{i'}$,
there is enough time to commit/abort all these transactions in $I_{z+1}$.
Hence, property (ii) holds.

\end{proof}




\begin{theorem}[FDS stability]
    \label{theorem:transaction-schedule-alg3}
In Algorithm \ref{alg:fully-distributed-scheduler}, for generation rate 
$\rho \leq \frac{1}{c_1d \log^2 s} \cdot \max\{ \frac{1}{k}, \frac{1}{\sqrt{s}} \}$,
where $d$ is worst distance of any transaction to the shards it will access and $c_1$ is some positive constant with burstiness $b \geq 1$, the number of pending transactions at any given round is at most $4bs$, and the transaction latency is at most  $2 \cdot c_1bd  \log^2 s\cdot min \{k, \lceil \sqrt{s}\rceil\}$.
\end{theorem}
\begin{proof}
    To estimate the number of pending transactions during a round, consider a round within a rescheduling period $I_i$. 
    From the proof of Lemma \ref{lemma:upper-bound-alg3}, the maximum number of old transactions during any round of $I_i$ is $2bs$. Within $I_i$, there can be at most $2bs$ newly generated transactions. Therefore, the upper bound on pending transactions during a round is $4bs$.

    For estimating transaction latency, we rely on the fact from Lemma \ref{lemma:upper-bound-alg3} that a transaction generated in a  period $I_z$ will be processed by the end of the next period. 
    Consequently, the transaction latency is bounded by twice the duration of the maximum interval length.
    From Lemma \ref{lemma:upper-bound-alg3}, rescheduling period length is $2\tau_{i'} = 60 H_2 b d_{i'} \cdot min \{k, \lceil \sqrt{s}\rceil\}$. We can replace $H_2= c \log s$ and $d_{i'}=c' d\log s$. We also combine positive constants $c$ and $c'$ as $c_1$ 
    which results in a latency of at most 
    $2 \cdot c_1 bd \log^2 s \cdot min \{k, \lceil \sqrt{s}\rceil\} 
    $.
\end{proof}

\section{Simulation Results}
\label{sec:simulation-result}
In this section, we provide our simulation results, conducted on a MacBook Pro with an Apple M1 chip featuring a 10-core CPU, a 16-core GPU, and 32 GB of RAM. The simulation was implemented using Python programming language with relevant libraries and resources. We considered the following specific parameters for simulation: the total number of shards ($s$) was set to $64$, the total number of accounts in the system was also set to $64$, considering one account per shard, and the maximum number of shards accessed by each transaction ($k$) was limited to $8$. The simulation used different values of the transaction generation rate ($\rho$) and burstiness ($b$). Each combination of these parameters was tested for $25000$ rounds.

\begin{figure*}[!ht]
\centering
\includegraphics[width=0.9\textwidth]{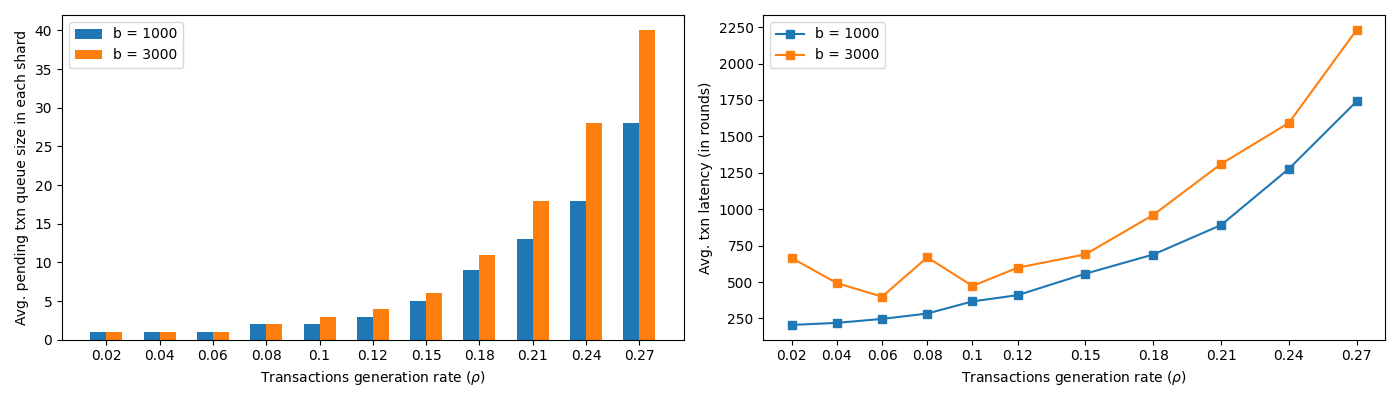}
\caption{Simulation results for Algorithm \ref{alg:basic-distributed-scheduler}: On the left, the average number of pending transactions in the pending queue of each home shard is shown versus $\rho$. On the right, the average transaction latency measured in rounds is plotted against $\rho$.}
\label{fig:simulation-result-alg1}
\end{figure*}

\begin{figure*}[!ht]
\centering
\includegraphics[width=0.9\textwidth]{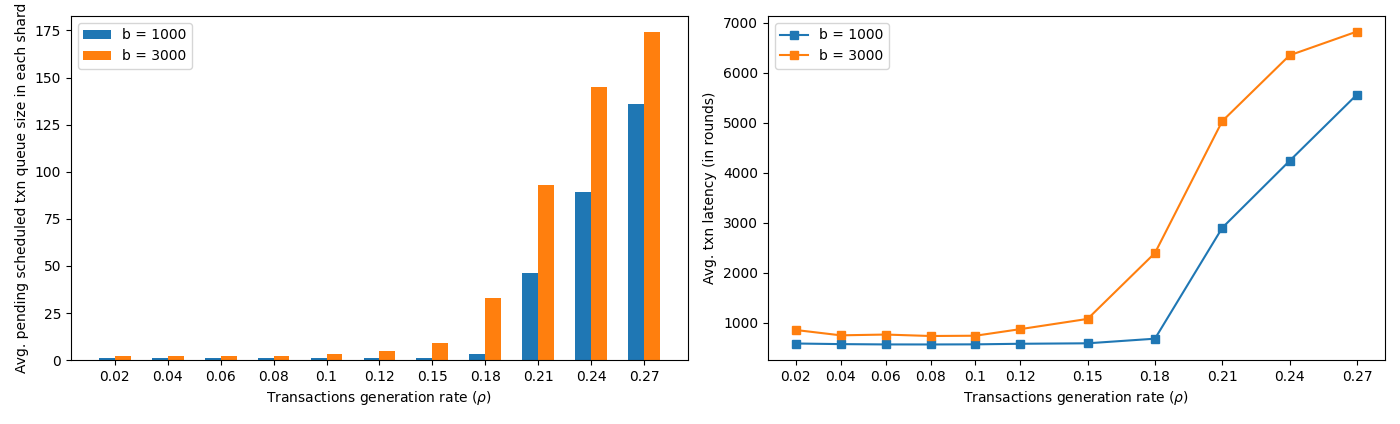}
\caption{Simulation results for Algorithm \ref{alg:fully-distributed-scheduler}: On the left, the average number of pending scheduled transactions in the queue (scheduled but not committed) of cluster leader shard is shown versus varying values of $\rho$. The average transaction latency measured in rounds is plotted against $\rho$ on the right.}
\label{fig:simulation-result-alg3}
\end{figure*}

 
In the simulation process, initially, we generated random, unique accounts and assigned them randomly to different shards, ensuring that each shard maintained its unique set of accounts. The simulation operated in epochs and continued until the defined total rounds were reached. Within each epoch, transactions were generated based on the previous epoch's length and the rate of transaction generation ($\rho$). Burstiness was introduced within only one epoch throughout the total rounds because the number of all possible adversarial strategies is over-exponential. Therefore, in simulations, we decided to focus on ``pessimistic'' strategies used in the adversarial queuing literature for proving impossibility results. They represent pessimistic scenarios where queues start being already loaded (burstiness) and in the remaining time, the system tries to prevent their further growth under the regular arrival of other transactions. All generated transactions were assigned randomly to different shards, and each shard maintains a pending transactions queue, sorted by transaction ID. For simplicity, we used a simple greedy coloring algorithm to color the transactions, which ensures conflicting transactions get different~colors. 


\paragraph{\bf Simulation Results of Algorithm \ref{alg:basic-distributed-scheduler}}

In this simulation, we consider the uniform model where each shard can send transaction information to any shard within the $1$ round.
Figure \ref{fig:simulation-result-alg1} shows the simulation results for Algorithm \ref{alg:basic-distributed-scheduler}. The left bar chart depicts the average pending transactions queue in each home shard as a function of the transaction generation rate ($\rho$) and burstiness ($b$). As expected, after the certain threshold of $\rho$, both the average pending transaction queue size and transaction latency increase exponentially. The queue size and transaction latency grow exponentially when $\rho > 0.15$. Moreover, the right line graph in Figure \ref{fig:simulation-result-alg1} shows the average latency of transactions in terms of rounds across various transaction generation rates. It is observed from the graph that transaction latency increases with increasing values of $\rho$ and $b$. We observed the average latency of transactions for values up to $\rho=0.15$ is under $750$ rounds.

\paragraph{{\bf Simulation Results of Algorithm \ref{alg:fully-distributed-scheduler}}}

We simulate Algorithm \ref{alg:fully-distributed-scheduler} by arranging shards on a line, where the distances between a pair of shards are dictated by their position in the line. We used 64 shards, $S_1$ to $S_{64}$. The distance between two adjacent shards in the line is set to $1$, indicating that the distance between $S_1$ and $S_2$ is $1$, and so forth. However, the distance from $S_1$ to $S_3$ is $2$, $S_1$ to $S_4$ is $3$, and so on. To organize the shards, we cluster them into layers and sublayers. The sublayers in each layer are constructed by shifting half of the diameter of that layer to the right.
In the lowest layer $l_i$, clusters consist of two shards each. Moving to the next higher layer, $l_{i+1}$, clusters now consist of four shards, and this pattern continues. In the highest layer, all shards are part of a single cluster. Each cluster has a designated leader $S_{\text{ldr}}$ responsible for coloring and scheduling all transactions within that cluster.

Figure \ref{fig:simulation-result-alg3} shows the simulation results of Algorithm \ref{alg:fully-distributed-scheduler}. The left bar chart shows the average pending scheduled (i.e. scheduled but not committed) transaction queue size in cluster leader shard as a function of the transaction generation rate ($\rho$), while the right line graph shows the average transaction latency. Particularly, the queue size does not exhibit exponential growth up to $\rho = 0.18$, and the average transaction latency is under $1000$ rounds for $\rho$ values from $0$ to $0.18$. Specifically, for $b=3000$ and $\rho=0.27$, the maximum average pending transactions are approximately $175$, and transaction latency reaches around $7000$ rounds. This value is significantly higher than the pending transactions and latency observed in Algorithm \ref{alg:basic-distributed-scheduler}, where the pending transactions are around $40$, and transaction latency is approximately $2250$ rounds. We observed that the queue size and transaction latency of Algorithm \ref{alg:fully-distributed-scheduler} grew significantly more than those of Algorithm \ref{alg:basic-distributed-scheduler}. Transaction latency of Algorithm \ref{alg:fully-distributed-scheduler} is higher than Algorithm \ref{alg:basic-distributed-scheduler} due to the non-uniform model and the distance between shards ranging from $1$ to $64$. The propagation and commitment of transactions take more time in this scenario compared to the uniform communication model of Algorithm \ref{alg:basic-distributed-scheduler}, where shards communicate in $1$ round.

\section{Discussion and Conclusions}
\label{sec:conclusion}

In conclusion, we provided an absolute upper bound on the stable transaction generation rate for a sharded blockchain system. Moreover, we design stable distributed scheduling algorithms for uniform and non-uniform sharded blockchain systems with bounded queue size and transaction latency. We also provide the simulation results of the proposed algorithms.
The blockchain system designers can use our results to design stable blockchain systems,
which are resilient to attacks that could overload the system with transactions.

We would like to note that for the basic scheduling algorithm (BDS) \ref{alg:basic-distributed-scheduler}, it is possible to use a deterministic distributed coloring algorithm \cite{ghaffari2022deterministic}. But this still requires to learn the degree of the conflict graph $\Delta$ and the total number of transactions.

For future work, one possible extension could be to study an asynchronous communication model for the proposed scheduler. Another could be to study the efficient communication mechanism between shards that have reduced message sizes.


\begin{acks}
This paper is supported by NSF grant CNS-2131538.
\end{acks}

\bibliographystyle{ACM-Reference-Format}
\balance
\bibliography{sample-base}

\end{document}